\documentclass[prl, reprint, twocolumn, floatfix]{revtex4}
\usepackage{graphics,epsfig,color,subfigure}
\usepackage{amssymb}

\makeatletter
\renewcommand{\@makecaption}[2]{
  \vskip\abovecaptionskip
  \sbox\@tempboxa{\small\sf #1: #2}%
  \ifdim \wd\@tempboxa >\hsize
  \small\sf #1: #2\par
  \else
    \global \@minipagefalse
    \hb@xt@\hsize{\hfil\box\@tempboxa\hfil}%
  \fi
  \vskip\belowcaptionskip}
\makeatother

\def\be{\begin{eqnarray}}
\def\ee{\end{eqnarray}}

\newcommand{\eqn}[1]{(\ref{#1})}

\def\Dslash{\,\,{\raise.15ex\hbox{/}\mkern-12mu D}}
\def\Dbarslash{\,\,{\raise.15ex\hbox{/}\mkern-12mu {\bar D}}}
\def\delslash{\,\,{\raise.15ex\hbox{/}\mkern-9mu \partial}}
\def\delbarslash{\,\,{\raise.15ex\hbox{/}\mkern-9mu {\bar\partial}}}
\def\pslash{\,\,{\raise.15ex\hbox{/}\mkern-9mu p}}
\def\calDslash{\,\,{\raise.15ex\hbox{/}\mkern-12mu {\cal D}}}

\def\lae{\mathrel{\mathop{\smash{\lower .5 ex \hbox{$\stackrel<\sim$}}}}}
\def\lae{\mathrel{\mathop{\smash{\lower .5 ex \hbox{$\stackrel>\sim$}}}}}

\begin{document}

\title{Universal Charge Diffusion and the Butterfly Effect}

\author{Mike Blake}

\affiliation{Department of Applied Mathematics and Theoretical Physics,
University of Cambridge, Cambridge CB3 0WA, United Kingdom}

\begin{abstract}
We study charge diffusion in holographic scaling theories with a particle-hole symmetry. We show that these theories have
 a universal regime in which the diffusion constant is given by $D_c = C v_B^2/ (2 \pi T)$ where  $v_B$ is the velocity of the butterfly effect.  The constant 
of proportionality, $C$, depends only on the scaling exponents of the infra-red theory. Our results suggest an unexpected connection between
transport at strong coupling and quantum chaos.
\end{abstract}


\maketitle

\paragraph{Introduction.} The transport properties of strongly correlated materials display a remarkable degree of universality. In particular many materials with fundamentally different microscopic physics exhibit a linear
resistivity over a broad temperature regime. A long-standing idea to explain such universality has been that transport is governed by a fundamental dissipative timescale $\tau \sim \hbar/(k_B T)$ \cite{subirbook, sachdev, planckian}. Recently, theoretical attention has been refocused on this claim following the direct observation of this `Planckian' timescale in a wide range of materials \cite{mackenzie,incoherent}. 
\paragraph{}Perhaps the most famous example of how $\tau$ could lead to universal behaviour is found in the proposed viscosity bound of Kovtun, Starinets and Son \cite{kss2}. Assuming that the viscosity, $\eta$, of a relativistic theory is controlled by this timescale leads to a conjecture that $\eta/s$ should be given by  
\be
\frac{\eta}{s} \sim \frac{1}{4 \pi}\frac{\hbar}{k_B}
\label{kss}
\ee
where $s$ is the entropy density. The original evidence for this claim stemmed from the observation that such a value is generic to many holographic theories with a gravity dual \cite{policastro,kss1,iqballiu}. However more recent developments in holography have shown that this relationship can receive large corrections, for instance in anisotropic systems or those without translational symmetry \cite{inhomo1,inhomo2,trivedi1,trivedi2,jorgeeta,matteoeta,leideneta}.
\paragraph{}Nevertheless, the idea that $\tau$ underpins the transport coefficients of strongly coupled matter has survived. Indeed, noticing that the viscosity controls momentum diffusion, it was proposed in \cite{incoherent} that one can reformulate the KSS bound in terms of the diffusion constants
\be
D \sim \frac{\hbar v^2}{k_B T} 
\label{diff}
\ee
where $v$ is a characteristic velocity of the theory. 
\paragraph{} A simple place in which to test \eqn{diff} is in the context of a particle-hole symmetric theory. In this case the electrical current decouples from
momentum and one has a finite conductivity even in a translationally invariant theory. In particular, for a holographic CFT with an Einstein gravity dual then the charge diffusion constant indeed takes a universal
form \cite{kovtun}
\be
D_c
 =  \frac{\hbar c^2}{4 \pi k_B T} \frac{{d+1}}{{d-1}} 
\label{cft}
\ee
where $d$ is the number of spatial dimensions and the characteristic velocity of a relativistic theory is the speed of light $v=c$. Outside the framework of a relativistic theory, however, it has proved challenging to identify a velocity to appear in \eqn{diff}.
\paragraph{}The purpose of this letter is to point out that a natural candidate for such a velocity in a strongly coupled theory is provided by the butterfly effect \cite{scrambling, butterflyeffect, stringyeffects, localisedshocks, multiple, chaos}.  In particular, the chaotic properties of strongly interacting large $N$ gauge theories have recently been intensely studied using the holographic correspondence \cite{scrambling,butterflyeffect, chaos, stringyeffects, localisedshocks, multiple,tangarife, kitaev,polchinski, roberts}. In such theories the butterfly effect refers to the exponential growth in the commutators of generic Hermitian operators which occurs after the thermal timescale $\beta$ \cite{butterflyeffect, localisedshocks, stringyeffects}%
\begin{eqnarray}
\langle  [\hat{W}_x(t_w), \hat{V}_y(0) ]^2 \rangle_{\beta} \sim f_1 e^{ \lambda_L(t_w - t_*- |x-y|/v_B)} +\dots
\label{chaoseqn2}
\end{eqnarray}
where $t_*$ is the scrambling time, $\lambda_L$ is the Lyapunov exponent and $v_B$ is known as the butterfly velocity.
\paragraph{}  For any holographic theory with a classical gravity dual both the Lyapunov exponent $\lambda_L$ and the butterfly velocity $v_B$ can be extracted from properties of a black hole horizon \cite{butterflyeffect, stringyeffects, localisedshocks}. 
Similarly, it has long been established that the DC transport coefficients of conserved quantities can also be related to the horizon via the membrane paradigm \cite{iqballiu}. As such, the holographic correspondence hints at an intimate connection between transport and the butterfly effect.
\paragraph{} Moreover the butterfly velocity  $v_B$ provides a natural analogue of the speed light  in \eqn{cft} which can be defined even in non-relativistic theories. Specifically, the commutator in \eqn{chaoseqn2} determines how a perturbation to the system by $\hat{V_y}$ propagates to affect a later, distant measurement by $\hat{W_x}$. The butterfly velocity therefore describes the finite speed at which information spreads, and hence has been argued in \cite{roberts} to act as a state-dependent Lieb-Robinson velocity \cite{lieb}. These general considerations therefore motivate us to propose $v_B$ as the characteristic velocity through which to formulate the diffusion bound of Hartnoll \cite{incoherent}. That is the diffusion constants should be bounded by 
\be
D \sim \frac{\hbar v_B^2}{k_B T}
\label{diffconjecture}
\ee
where the saturation of such a bound would correspond to a `Planckian' dissipation time $\tau \sim \hbar/ k_B T$.
\paragraph{} In the remainder of this letter we wish to provide initial evidence in support of \eqn{diffconjecture} by studying the charge diffusion constant of simple holographic scaling geometries.
In particular we will consider theories whose infra-red physics is described by a dynamical critical exponent, $z$, a  hyper-scaling violation exponent, $\theta$, and an anomalous dimension $\Phi$ for the charge density. Since previous studies of the butterfly effect in holography have focused on conformal field theories \cite{butterflyeffect, stringyeffects, localisedshocks}, our first task is to calculate the velocity $v_B$ for these more general geometries.
\paragraph{} Armed with this velocity, we can then proceed to compare with the diffusion constant. Our central result is that in these theories there is a universal regime in which $D_{c}$ indeed satisfies a relationship of the form \eqn{diffconjecture} 
\be
D_{c} = \frac{d_{\theta}}{\Delta_{\chi}} \frac{ v_B^2}{2 \pi T} 
\label{resultintro}
\ee
where $d_{\theta}$ is the effective spatial dimensionality of the fixed point, $\Delta_{\chi}$ is the scaling dimension of the susceptibility and we have reverted to high energy units $\hbar = k_B=1$. Finally we close this letter with a brief discussion of the implications of \eqn{resultintro} for our general proposal, and the extension to other diffusion constants. 
\paragraph{The butterfly effect in scaling geometries.} Like in classical physics, the butterfly effect in a quantum system is associated with whether the effects of a small perturbation can eventually become large at late times.  
In holographic theories, the butterfly effect corresponds to the fact that the energy of an in-falling particle near a black hole horizon is exponentially boosted at late times \cite{butterflyeffect}. The back-reaction of this particle on the geometry creates a shock-wave along the horizon that causes the growth of commutators \eqn{chaoseqn2}.
\paragraph{} From the form of this shock-wave geometry, it is possible to read off both the Lyapunov exponent and the velocity $v_B$. For the most part, previous holographic studies of chaos have focused on conformal field theories in which the velocity $v_B$ is just a constant \cite{butterflyeffect, stringyeffects, localisedshocks} . We will therefore begin by adapting the shock-wave techniques of \cite{butterflyeffect, stringyeffects, localisedshocks} to calculate $v_B$ for a more general family of metrics. These are described by an infra-red geometry 
\be
ds_{d+2}^2 = - U(r)dt^2 + \frac{dr^2}{U(r)} + V(r) d \vec{x_d}^2
\label{metric}
\ee
where $r$ is the additional radial coordinate of the gravitational theory.  At zero temperature we will assume we have power law solutions
\be
U(r) = L_t^{-2} r^{u_1}  \;\;\; V(r) =L_x^{-2} r^{2 v_1}
\label{hyperscalingmetric}
\ee
where we take $u_1 > 1, v_1> 0$ so that $r \rightarrow 0$ corresponds to the infra-red of our theory.  To turn on a finite temperature we can introduce a horizon at $r = r_0$
\be
U(r) = L_t^{-2} r^{u_1} \bigg( 1 - \frac{r_0^{\delta}}{r^{\delta}}\bigg) \;\;\;\;\; V(r) = L_x^{-2} r^{2 v_1}
\label{powerlawmetric}
\ee
with $\delta = d v_1 + u_1 - 1$. The temperature of the boundary quantum field theory is then related to the horizon radius $r_0$ by the usual formula $4 \pi T = U'(r_0) $.
\paragraph{}Although written in unusual coordinates, these metrics simply correspond to a family of hyper-scaling violating geometries \cite{strangemetals, huijse, dong, kachru,elias1,elias2, pang} 
where the critical exponents $z, \theta$ are related to the power laws in the metric via
\be
u_1 = \frac{2 z - 2 \theta/d}{z - 2 \theta/d} \;\;\;\ 2 v_1 = \frac{2 - 2 \theta/d}{z - 2 \theta/d}
\ee
As usual the dynamical critical exponent, $z$, characterises the different scaling of space and time at the fixed point $
[x] =-1$, $ [T] = -[t] = z $. The hyperscaling violation exponent $\theta$ corresponds to the fact that the metric transforms non-trivially under scaling 
and is responsible for an effective shift in the dimensionality of the free energy $
 [f] = z  + d - \theta = z + d_{\theta}$. 
\paragraph{}Additionally we have retained various parameters in our solution, $L_t, L_x$ that might normally be set to unity. These parameters, which set various scales in our infra-red theory, are non-universal and will depend on the embedding of our metric into an asymptotically Anti-de Sitter (AdS) space-time. We will retain them in order to emphasise that the relationship between the diffusion constant and the butterfly effect \eqn{resultintro} is independent of this ultraviolet (UV) data.  
\paragraph{Shock wave geometries.} We now wish to calculate the velocity of the butterfly effect dual to the metrics \eqn{powerlawmetric} by constructing the relevant shock wave geometries \cite{localisedshocks, butterflyeffect}. In order to do this, we first need to pass to Kruskal coordinates $(u,v)$. We therefore define
\be
u v = -e^{U'(r_0) r_{*}(r)} \;\;\;\;\;\;\;\; u/v = -e^{-U'(r_0) t}
\ee
where the tortoise coordinate is given as usual by $dr_{*} = dr/U(r)$. In terms of these coordinates our black hole metric now reads
\begin{eqnarray}
ds^2 &=&  A( u v) du dv + V( u v) d\vec{x_d}^2 \nonumber \\
A( u v) &=& \frac{4}{u v}\frac{U(r)}{U'(r_0)^2} \;\;\; V( uv) = V(r) 
\end{eqnarray}
with horizons now located at $u=0$ and $v=0$. 
\paragraph{} To study the butterfly effect, we consider releasing a particle from $x=0$ on the boundary of AdS at a time $t_w$ in the past. Then for late times (i.e. $t_w > \beta$) the energy density of this particle in Kruskal coordinates is exponentially boosted and localised on the $u=0$ horizon 
\be
\delta T_{u u} \sim E e^{\frac{2 \pi}{\beta} t_w} \delta(u) \delta(\vec{x})
\label{stresstensor0}
\ee
where $E$ is the initial asymptotic energy of the particle. As a result, even the effects of an initially small perturbation cannot be neglected and after the scrambling time $t_*  \sim \beta \; \mathrm{log}\; N^2$ the back-reaction of the stress tensor \eqn{stresstensor0} on the metric becomes significant. 
\paragraph{}Within generic theories of Einstein gravity coupled to matter the resulting geometry takes a universal form - it is a shock-wave that is localised on the horizon \cite{butterflyeffect,stringyeffects, localisedshocks,camanho,sfetsos,dray}. In particular such a solution corresponds to a shift in the $v$ coordinate $v \rightarrow v + h(x)$ as one crosses the $u=0$ horizon. The resulting metric can then be written as
\be
ds^2 = A( u v) du dv +  V( u v ) d\vec{x}^2 - A(u v) \delta(u) h(x) du^2
\label{metricshift}
\ee
where one finds a solution to the Einstein equations provided the shift obeys \cite{sfetsos}
\be
(\partial_{i}\partial_{i}  - m^2 ) h(x) \sim \frac{1 6 \pi G_N V(0)}{A(0)} E e^{\frac{2 \pi}{\beta} t_w} \delta(\vec{x})
\label{poisson}
\ee
and the screening length $m$ is given by
\be
m^2 = \frac{d}{A(0)} \frac{\partial{V(u v)}}{\partial{ (u v) }}\bigg|_{u=0}
\label{musquare}
\ee
Remarkably, after using the background equations of motion, one finds that the equations \eqn{poisson} and \eqn{musquare} for the shift still hold even when there is a non-trivial stress tensor
supporting the background geometry \cite{sfetsos, roberts}. 
The only way the matter content of the theory effects the shock-wave is indirectly through the determination of the  metric functions $A(u v)$ and $V(u v)$. 
\paragraph{}The net result is that all we need to do to study the butterfly effect is therefore to solve \eqn{poisson}.  At long distances $x \gg m^{-1}$ the metric is simply given by
\be
h(x) \sim \frac{E  e^{\frac{2 \pi}{\beta} (t_w -t_{*})- m |x|}}{|x|^{\frac{d-1}{2}}}
\ee
Since it is the formation of this shock wave geometry that is responsible for the growth of commutators \cite{butterflyeffect,localisedshocks}  
 one can immediately read off the Lyapunov exponent $\lambda_L$ and velocity $v_B$ of these holographic theories as 
\be
\lambda_L = \frac{2 \pi}{\beta} \;\;\;\;\;\; v_B = \frac{2 \pi}{\beta m}
\ee
\paragraph{}Whilst the Lyapunov exponent is universal in all these theories, and saturates the proposed bound on chaos \cite{chaos}, the velocity
is model dependent.  In order to extract $v_B$ for our metrics \eqn{powerlawmetric} we observe that we can rewrite the
screening length \eqn{musquare} in terms of more familiar coordinates $(r,t)$ as
\be
m^2 = d \pi T V'(r_0) 
\ee
where the $'$ indicates a radial derivative. The butterfly velocity therefore takes a remarkably simple form
\be
v_B^2 = \frac{4 \pi T}{d V'(r_0)} 
\label{vbsquare}
\ee
For the case of the AdS-Schwarzchild solution, dual to a CFT, we have $V'(r_0) \sim  r_0 \sim T$ giving a constant velocity 
as expected. In our more general scaling geometries, however, this velocity has a non-trivial temperature dependence 
\begin{eqnarray}
v_B^2 \sim T^{2 - 2/z}
\label{velocityscaling}
\end{eqnarray}
\paragraph{Charge diffusion.}Now that we have the butterfly velocity, $v_B$, we can proceed to study the diffusion of a $U(1)$ charge in the background \eqn{powerlawmetric}. 
We therefore consider coupling a gauge field $A_{\mu}$ to our theories using an action
\be
S = \int \mathrm{d}^{d+2}x \sqrt{-g} \bigg[ -\frac{1}{4} Z(r) F^{\mu \nu} F_{\mu \nu} \bigg]
\label{maxwell}
\ee
where $Z(r)$ is a position dependent Maxwell coupling. In the context of our scaling geometries, this position dependence can arise from a coupling between the
gauge field and a logarithmically running dilaton that supports our background geometry \eqn{powerlawmetric}. We will therefore assume that it takes a power law form
$ Z(r) = Z_0 r^{\gamma}$. 
\paragraph{} In terms of the boundary field theory, this running of the Maxwell coupling corresponds to the possibility that the $U(1)$ charge density can have an anomalous dimension in the IR \cite{blaise1,blaise2,andreas1,andreas2}. That is the chemical potential, $\mu$, and charge density, $\rho$, dual to the gauge field $A_{\mu}$ have scaling dimensions
\be
[\mu] = z - \Phi  \;\;\;\;\;\;\;\;\;\;\  [\rho] = d - \theta + \Phi
\ee
where the anomalous dimension $\Phi$ is related to the running of the Maxwell coupling by
\be
\gamma  =\frac{2 \Phi - 2 \theta/d}{z - 2 \theta/d}
\ee
\paragraph{}Extracting the diffusion constant for a current with particle-hole symmetry is then a standard calculation in holography. The Einstein relation 
$ D_{c} = \sigma/\chi $ relates the diffusion constant to the conductivity $\sigma$ and the susceptibility $\chi = (\partial \rho/ {\partial \mu})_T$. Both of these quantities can then be 
calculated using the membrane paradigm \cite{iqballiu}. The electrical conductivity takes a particularly simple form - it is just related to the effective Maxwell coupling on the horizon:
\be
\sigma  = V^{d/2-1} Z(r) \bigg|_{r_0}  
\label{dcconductivity}
\ee
\paragraph{}Obtaining the susceptibility is only slightly more complicated. If we consider turning on a small chemical potential, $\mu$, then the Maxwell equation implies that the electric flux is a constant
\be
-\sqrt{-g} Z(r) g^{rr} g^{tt} \partial_{r} A_t = \rho
\ee
where $\rho$ is interpreted as the charge density of the boundary theory.  Since the metric functions are independent of $\mu$ at leading order we can read off the susceptibility
\be
\chi^{-1} = \frac{\partial {\mu}}{\partial {\rho}}\bigg|_{\rho=0} = \int_{\infty}^{r_0} \mathrm{d}r \frac{1}{\sqrt{-g} Z(r) g^{rr} g^{tt}}
\label{suscint}
\ee
and hence arrive at the diffusion constant 
\be
D_{c} =   [V^{d/2-1} Z(r)]_{r_0}  \int_{\infty}^{r_0} \mathrm{d}r \frac{1}{\sqrt{-g} Z(r) g^{rr} g^{tt}}
\label{diffint}
\ee
At this point it is important to emphasise that in general the the susceptibility, and hence diffusion constant, depends via \eqn{suscint}
on the details of the full bulk geometry. Since the butterfly velocity depends only on the local properties of the horizon, these effects will not
always be related in a simple manner. 
\paragraph{}In particular, the behaviour of the susceptibility will depend on whichever region of the geometry dominates the
integral in \eqn{suscint}. For our scaling geometries \eqn{powerlawmetric} this results in two qualitatively different regimes, 
depending on the scaling dimension  
\be
\Delta_{\chi} = [\rho] - [\mu] = d_{\theta} + 2 \Phi - z 
\ee
of the susceptibility.
\paragraph{Non-universal regime.}
When the scaling dimension is negative, $\Delta_{\chi}/z < 0$, then it is the UV region of the geometry that controls the susceptibility. Since in this case the diffusion constant is sensitive to the full geometry, we should not expect to find universal behaviour in general. In particular after performing the integral \eqn{suscint} we find that at low temperatures the susceptibility takes a constant value set by the cut-off $\chi \sim \Lambda_{UV}^{\Delta_{\chi}/z}$. 
The resulting diffusion constant therefore scales as
\be
D_{c}  \sim \bigg(\frac{\Lambda_{UV}}{T} \bigg)^{-\Delta_{\chi}/z}T^{1- 2/z}
\ee
and hence in this regime it is parametrically larger than $v_B^2/T$ \eqn{vbsquare} by powers of the cut-off. 
\paragraph{Universal regime.} In contrast, when the scaling dimension $\Delta_{\chi}/z >0 $ it is the infra-red region of the geometry that dominates the integral and gives 
a susceptibility $\chi \sim T^{ \Delta_{\chi}/z}$. As a result, the charge fluctuations are now controlled by the near-horizon geometry and so we are able to relate $D_c$
to the butterfly effect. Indeed, if we explicitly evaluate our expression \eqn{diffint} then we arrive at a formula for the diffusion constant
\be
D_{c} =  \frac{z -2 \theta/d}{\Delta_{\chi}} {L_x^2} r_0^{1 - 2 v_1}
\ee
where the explicit dependence on the dilaton profile has cancelled between the conductivity and susceptibility. If we now note that our formula \eqn{vbsquare} for the butterfly velocity is equivalent to
\be
\frac{v_B^2}{2 \pi T} = \frac{z -2 \theta/d}{d_\theta} {L_x^2} r_0^{1 - 2 v_1}
\ee
then we see that the diffusion constant always takes the form
\be
D_{c}  = \frac{d_\theta}{\Delta_{\chi}} \frac{v_B^2}{2 \pi T}
\label{result}
\ee
which is the central result we presented in the introduction \eqn{resultintro}. 
\paragraph{} It is worth emphasising that both the diffusion constant $D_{c}$ and the butterfly velocity $v_{B}$ depend, through $L_x$ and $r_0$, on the details of our holographic
theory. In particular they are sensitive to how our metric \eqn{powerlawmetric} is embedded into an asymptotically AdS space-time. The relationship between them, on the other hand,
is completely universal -  it depends only on the scaling exponents of the infra-red fixed point. 
\paragraph{Discussion.} In summary we have proposed that a natural way to define the diffusion bound of Hartnoll \cite{incoherent} in a strongly coupled system is to use the velocity, $v_B$, of the butterfly
effect. Our main piece of evidence in support of this proposal was a calculation of the charge diffusion constant of simple holographic scaling geometries. We found that these theories had
a universal regime in which the diffusion constant could be tied to properties of the black hole horizon. Since the butterfly velocity is also determined by the horizon, we were able to establish
our central result \eqn{result}. The net result is that we have provided an intuitive picture of charge diffusion in these models. That is in this regime we can think of charge transport as being governed 
by a diffusive process with a velocity $v_B$ and a Planckian timescale $\tau \sim 1/T$. 
\paragraph{} More generally we saw that the diffusion constant will not take a universal value but rather depends on the details of the full geometry. Since these effects increased
 the diffusion constant, this is consistent with the proposal that \eqn{diffconjecture} corresponds to a `lowest possible value' for $D_{c}$. However, it is worth noting that it does not seem 
possible to formulate a strict bound, at least in terms of $v_B$, since the coefficients in \eqn{result} depend on the universality class.  Nevertheless, in generic theories these are order one numbers 
and so the essential point is that as $T \rightarrow 0$ the diffusion constant will be at least as big as $v_B^2/T$.
\paragraph{} Whilst our focus in this letter has been on studying charge diffusion in hyperscaling violating geometries, it is clearly of interest to understand to what extent 
our proposal \eqn{diffconjecture} holds more generally. One obvious direction to pursue would be to investigate how higher derivative corrections to the bulk action affect
our result \eqn{result}. Additionally it is important to extend our analysis to include other diffusion constants. In the Appendix we therefore calculate 
the momentum diffusion constant (i.e. viscosity) of these holographic scaling geometries and demonstrate that it is also consistent with \eqn{diffconjecture}.
\paragraph{}Finally, in \cite{mikeincoherent} we will provide a further test of our proposal by studying more general holographic models with momentum
relaxation \cite{lattices, univdc, andrade, donos1, donos2, saso1, saso2,blaise2}. This will also allow us to study the energy diffusion constant, which diverges in the translationally invariant theories considered here. 
Remarkably we find that when momentum relaxation is strong the diffusion constants do not become arbitrarily small, but rather are universally given by \eqn{diffconjecture}.  As such these models constitute strong further evidence that the diffusion constants of holographic theories are indeed bounded in terms of $v_B$.  
\paragraph{Note Added} After this work appeared we became aware that Dan Roberts and Brian Swingle have simultaneously computed the velocity of the butterfly effect in hyper-scaling violating geometries \cite{roberts}. Their formula agrees with our expression \eqn{vbsquare}.  
\paragraph{Acknowledgements}  I  thank Dan Roberts, Brian Swingle and Aristomenis Donos for useful conversations, and am particularly grateful to Jorge Santos for countless discussions about the butterfly effect. I also thank Aristomenis Donos, Sean Hartnoll, Jorge Santos and David Tong for their comments on a draft of this paper. I am also grateful to the anonymous referees whose comments improved the presentation and discussion of these results. I am funded through a Junior Research Fellowship from Churchill College, University of Cambridge. This work was supported in part by the European Research Council under the European Unions Seventh Framework Programme (FP7/2007-2013), ERC Grant agreement STG 279943, Strongly Coupled Systems. 
%


\appendix{}
\section{Appendix: Shear Viscosity}

\paragraph{} As we mentioned in the main text, in relativistic theories the shear viscosity $\eta$ is equivalent to the momentum diffusion constant \footnote{Note that if translational symmetry is broken as in \cite{jorgeeta,matteoeta,leideneta} then momentum no longer diffuses. As such we would not expect our analysis to apply to the viscosity in this case.}
\be
D_{p} = \frac{\eta }{\epsilon + {\cal P}} 
\ee
where $\epsilon$ is the energy density and ${\cal P}$ the pressure. In the absence of chemical potentials, one can extract the momentum diffusion constant of our isotropic scaling solutions from a Maxwell action with a position-dependent coupling $Z(r) = V(r)/(16 \pi G_N)$ \cite{kss1, iqballiu}. As such much of the analysis we presented for charge diffusion can be directly applied to the viscosity of these models. In particular we again have a universal regime where the shear diffusion constant is universally related to the butterfly effect
\be
D_{p} =  P \frac{v_B^2}{2 \pi T}
\ee
where the constant is now $ P = d_{\theta}/(d_{\theta} + 2 - z)$.  
\paragraph{}This behaviour in the shear diffusion constant is also relevant in the case of anisotropic theories. For concreteness we can consider a 5 dimensional bulk metric with anisotropy in the $z$ direction
\be
ds^2 = -U(r)dt^2 + \frac{dr^2}{U(r)} + g_{xx}(r)(dx^2 + dy^2) + g_{zz}(r) dz^2 \nonumber
\ee
For such an anisotropic metric it is well known that the longitudinal viscosity, $\eta_{xz}$, can deviate significantly from the KSS bound \cite{inhomo1,inhomo2,trivedi1,trivedi2} 
\be
\frac{\eta_{xz}}{\eta_{xy}} =  \frac{g_{xx}}{g_{zz}} \bigg|_{r=r_0} \;\;\;\;\;\;\;\;\;\;\;\;\;\ \eta_{xy} = \frac{s}{4 \pi}
\label{visc}
\ee
Since $\eta_{xz}$ can then be made arbitrarily small, one might be tempted to conclude that these anisotropic metrics violate a Planckian bound on $\tau$. 
\paragraph{}It is straightforward, however, to repeat our discussion of diffusion in these backgrounds. The two viscosities in \eqn{visc} are related to shear diffusion constants $D_z, D_y$ which measure how the momentum $p_x$ diffuses in the $z$ and $y$ directions
\be
D_{z} = \frac{\eta_{x z}}{\epsilon + {\cal P}_{\perp}} \;\;\;\;\;\; D_{y} = \frac{ \eta_{x y}} {\epsilon + {\cal P}_{\perp}}
\ee
with ${\cal P}_{\perp}$ the pressure in the transverse plane. In order to extract the timescales underlying these diffusion constants, we need to determine the characteristic velocities $v_B$ in the $y$ and $z$ directions. In an anisotropic theory these velocities are no longer necessarily the same. Indeed upon constructing the shock wave solution we find that they are related by precisely the same ratio of metric functions that appeared in the viscosities \eqn{visc}
\be
\frac{v_z^2}{v_y^2} = \frac{g_{xx}}{g_{zz}} \bigg|_{r=r_0}
\ee
The net result is that the difference in $D_z$ and $D_y$ precisely reflects the difference in these velocities. In particular if we assume power law behaviour in the metric functions then we again have a universal regime in which the diffusion constants are
\be
D_{{z}} \sim \frac{v_z^2}{2 \pi T} \;\;\;\;\ D_{y} \sim \frac{v_y^2}{2 \pi T}
\label{diffani}
\ee
We can therefore see that the fact that $\eta_{x z}$ (or equivalently $D_{z}$) can be made arbitrarily small does not imply that $\tau$ is becoming sub-Planckian. Rather, it corresponds to the fact that the characteristic velocity in the $z$ direction is suppressed by the anisotropy. That is, in the universal regime,  the timescale underlying both $D_z$ and $D_y$ will still always be given by $\tau \sim 1/T$. 

\end{document}